# Integrated structural and functional optical imaging combining spectral-domain optical coherence and multiphoton microscopy

Claudio Vinegoni, Tyler Ralston, Wei Tan, Wei Luo,

Daniel L. Marks, Stephen A. Boppart

Biophotonics Imaging Laboratory

Beckman Institute for Advanced Science and Technology

Department of Electrical and Computer Engineering

Department of Bioengineering

Colleges of Engineering and Medicine

University of Illinois at Urbana-Champaign

**Abstract:**

An integrated microscope that combines different optical techniques for simultaneous imaging is demonstrated. The microscope enables spectral-domain optical coherence microscopy based on optical backscatter, and multi-photon microscopy for the detection of two-photon fluorescence and second harmonic generation signals. The unique configuration of this integrated microscope allows for the simultaneous acquisition of both anatomical (structural) and functional imaging information with particular emphasis for applications in the fields of tissue engineering and cell biology. In addition, the contemporary analysis of the spectroscopic features can enhance contrast by differentiating among different tissue components.

**Main text:**

Optical coherence tomography (OCT) [1] has been demonstrated as a non-invasive and high resolution imaging technique for locating and diagnosing pathological tissue. However, only the linear characteristics of the tissue, such as scattering, birefringence, absorption, and refractive index variations contribute to producing image contrast. Moreover, these properties can be similar in different tissues making it very difficult, for example, to distinguish between normal and pathological tissues. Different approaches have been pursued in order to solve this problem and in particular, OCT techniques that exploit molecular differences like the use of exogenous contrast agents or techniques that exploit coherent nonlinear optical methods to identify endogenous molecular properties, have shown promise [2,3].

Recently, an alternative approach has been proposed that consists of combining different optical imaging techniques to provide different yet complementary information [4-6]. Multiphoton microscopy (MPM) has been shown to be a highly advantageous approach for biological imaging since its first exploitation [7]. The main advantages include three-dimensional (3D) optical sectioning from thick scattering samples by way of restricting the fluorophore excitation only to the focal plane, reducing the autofluorescence background, increasing the depth of imaging, and reducing cell and tissue damage by the use of longer near-infrared light [8]. Optical coherence microscopy (OCM) [9], extends the capabilities of OCT and confocal microscopy by combining high sensitivity and coherence-gated detection with confocal optical sectioning. The result is an improvement in the rejection of unwanted scattered light generated from points outside of the imaging plane [10]. A dramatic enhancement of the image contrast is therefore possible, at imaging depths in highly-scattering tissue that exceed confocal or multiphoton microscopy alone.

A microscope that combines both OCM and MPM imaging functions into a single instrument would enable a wide range of biological investigations, with MPM based on the detection of the fluorescence emitted by endogenous or exogenous markers and OCM delivering information on the endogenous scattering properties of the sample. These two modalities therefore provide different yet complementary imaging contrast mechanisms, increasing the information extracted from the sample.

Particularly useful applications for such an investigative tool can be found in a multitude of fields including tissue engineering, and cell, tumor, and plant biology. By overlaying the site-specific functional image from MPM (i.e. an excited fluorescent marker implies a functional protein) on the background structural image obtained from OCM, a more comprehensive view of different tissues can be obtained. It is important to note that such an instrument could also be easily used (without any modification) to perform laser ablation, where the absorption is still a two-photon process but at much higher incident pulse energies. This feature could be very useful in order to do high resolution optical ablation followed by optical histology obtained by OCM for basic investigative studies in cell and tissue responses, as well as for medical applications.

The experimental setup used in this work is shown in Fig. 1. The light source consists of a frequency-doubled Nd:YVO$_4$-pumped Ti:sapphire laser with a center wavelength of 800 nm, a bandwidth of 60 nm, an 80 MHz pulse repetition rate, and an average power on the sample that can be varied between 1 to 10 mW. Due to the highly dispersive nature of the microscope objective (20x, 0.9 NA, water-immersion, Olympus, Inc.), the pulses are first precompensated by using a double-fold prism path. The path length was initially fixed after obtaining the maximum two photon signal from a 100 µM solution of Rhodamine 6G dye sandwiched between two microscope cover slips. The beam is directed to a scan head that consists of two galvanometer-controlled mirrors for high-speed acquisition, and then beam-expanded in order to match the back-aperture of the objective in this upright microscope configuration. Samples were fixed on a holder attached to a 3D translation stage that allowed for large scanning areas (25mm x 25mm), a feature particularly useful for *in situ* imaging of animal models. A z-axis translation stage allowed for the acquisition of image stacks, and subsequently for 3D volume rendering of the data. The laser source serves both as an excitation source for two-photon absorption and as a low-coherence source for OCM, facilitating the multi-modality imaging capabilities of this microscope.

The photons resulting from the two photon absorption process are emitted over the entire 4π solid angle. A portion of the signal is collected in the backward direction by the objective and deflected by a dichroic beam-splitting mirror (Cold Mirror, CVI Laser, Livermore, CA). The photons are then collected and coupled into a multi-mode fiber. This modular configuration allows for multiple different detection units, if desired, without modification to the original setup. A short-pass filter (BG39, CVI Laser, Livermore, CA) located prior to the fiber is used to filter out scattered pump photons. The filter can be easily interchanged for use with different selected fluorophores or for second harmonic generation imaging. For this study, a photomultiplier tube (H7421-40, Hamamatsu, Inc.) working in photon-counting mode was used to detect two-photon fluorescence, with a maximum quantum efficiency of 30% at 580 nm, and a dark counts rate of 100 Hz. When overfilling the back-aperture (18 mm) of the objective, axial and lateral two-photon resolutions of 0.8 μm and 0.5 μm, respectively, were measured using fluorescent nanobeads.

The OCM detection scheme in this integrated microscope is different with respect to a previously reported system [4]. We have implemented a spectral-domain OCM system, instead of time-domain OCM system, with many distinct advantages. While standard OCM requires two scans to be performed (axial and lateral scanning) for the acquisition of 3D data, the spectral-domain technique can be implemented using only lateral scanning, as each spectral line-scan corresponds to optical signals over larger distances in depth through the sample. It has recently been shown that the spectral-domain configuration provides significant advantages in terms of acquisition speed, sensitivity, and simplicity in the acquisition module; benefits that are incorporated into our integrated microscope [11]. In practice, sensitivities of the order of more than 100 dB are commonly achieved with acquisition speed in the range of 30,000 spectral acquisitions per second. Moreover, spectral-domain OCM inherently provides direct access to the spectral information for spectroscopic OCM signal analysis [11], and the absence of any moving reference arm in the setup provides an inherent optical phase stability that makes this modality ideally suited for the evaluation of the spectral components in the interference pattern. In fact, because different tissue structures and molecules have different spectral absorption and scattering properties, the spectral analysis, combined with the coherence gating, can increase the OCM image contrast, with the potential for generating spatial maps of molecules within the sample. Finally, inverse scattering computations show that the out-of-focus data acquired in this spectral-domain OCM configuration can be significantly resolved, resulting in multi-plane acquisition [12].

In our setup (Fig. 1), light is collimated and dispersed off of a blazed diffraction grating having 830.3 grooves per millimeter. The optical spectrum is focused using a pair of achromatic lenses which have a combined focal length of 150 mm. The focused light is incident on a line-scan camera (L104k–2k, Basler, Inc.) which contains a 2048-element charge-coupled device array of detection elements. This camera has a maximum readout rate of 29 kHz, thus one axial scan (corresponding to one pixel in an *en face* optical section) can be captured during an exposure interval of 34 μs. Digital processing of the detected signal included a spline interpolation to make the signal more uniform, and a discrete Fourier transform on each set of 2048, 10-bit, values captured by the line-scan camera to transform the signal from the frequency (spectral) domain into the spatial (depth) domain. The scattering amplitudes corresponding to the focus in each adjacent axial scan were subsequently assembled into two-dimensional (2D) *en face* images for visualization on a personal computer. Acquisition and visualization of OCM and MPM images was performed simultaneously. We note that in our system, the confocal gating (confocal parameter = 2.2 µm) from the high numerical aperture objective was shorter in distance (higher axial resolution) than the coherence gating (coherence length = 4.7 µm) from the broad 60 nm spectral bandwidth of the laser source. A transverse resolution of 0.9 µm was measured from a calibrated U.S. Air Force Resolution test target using the edge-scan definition.

This integrated microscope has applications in a wide range of fields, some of which are presented here. Figure 2 shows structural and functional imaging of *in vitro* smooth muscle tissue from a transgenic green-fluorescent protein (GFP) mouse. OCM and MPM image data both strongly correlate with the corresponding histological data. While OCM and MPM images were acquired on the same area simultaneously, the histological image was acquired from a morphologically-matched area following histological processing of the tissue specimen.

In addition to the larger morphological features noted in the highly-scattering muscle, this system is particularly useful for obtaining high-resolution structural and functional information in applications that involve both biological and non-biological microscopic elements, such as in tissue engineering. Microstructural data of cell-scaffolds, as well as image data of cells, their shapes, and their positions, can be readily acquired. In addition, complementary functional information of cell dynamics and cell-cell and cell-scaffold interactions can be obtained, such as tracking nuclei and the expression of focal adhesion molecules such as vinculin. To demonstrate these capabilities, fibroblast cells transfected with GFP-labeled vinculin and stained with a nuclear dye (Hoechst 33342) were seeded and cultured on a microtextured poly(dimethyl-siloxane) elastomer substrate. The microtextured substrate consisted of 25 μm high and 25 μm wide microgrooves, separated by 100 μm wide ridges. Mechanical stimuli provided through surface textures (surface topographies) have been reported to influence cell shape, gene expression, protein production and deposition, cell proliferation, migration, differentiation, and survival. Therefore, exploring the effects of microscale textures at the cell-scaffold interface provides an attractive approach to enhance cell behavior without destabilizing the delicate biochemical condition. The OCM image in Fig. 3 demonstrates the structural information of the cell and scaffold, while the MPM images reveal the localization of the nucleus and the expression of the GFP-labeled vinculin. The combined OCM-MPM images provide complementary and additive information regarding the cell in relation to its local environment.

Another application involves the use of the integrated microscope to dynamically study cell-scaffold interactions under more physiological 3D conditions, as shown in Fig. 4. Fibroblast cells (as described above) were cultured in a 3D Matrigel matrix. The OCM image shows the 3D cell morphology and position as well as the fibrous network of the matrix. The MPM image shows the cell nuclei and the vinculin expression in the cells. Compared to the elongated oval-shaped fibroblast cells cultured in 2D cell culture (not shown here), cells in a 3D Matrigel matrix display a rounded morphology with spherical nuclei. Because the Matrigel matrix is composed of laminin and collagen IV and lacks many of the proteins associated with cell adhesion, vinculin is not as highly expressed, and the two-photon fluorescence resulting from interactions between the cells and the Matrigel matrix is therefore less.

In conclusion, we have developed and applied an integrated microscope that is capable of simultaneous image acquisition from multiple optical imaging modalities. This instrument provides a new investigational tool for the visualization of structure and function in fields such as tissue engineering and tumor biology.

This research was supported in part by the National Institutes of Health (1 R01 EB00108, S.A.B.). Additional information can be found at http://biophotonics.uiuc.edu.

# REFERENCES

[1] D. Huang, E.A. Swanson, C.P. Lin, J.S. Schuman, W.G. Stinson, W. Chang, M.R. Hee, T. Flotte, K. Gregory, C.A. Puliafito, and J.G. Fujimoto, Science 254, 1178 (1991).

[2] C. Vinegoni, J.S. Bredfeldt, D.L. Marks, and S.A. Boppart, Opt. Exp. 12, 331 (2004).

[3] Y. Jiang, I.V. Tomov, Y. Wang, and Z. Chen, App. Phys. Lett. 86, 133901 (2005).

[4] E. Beaurepaire, L. Moreaux, F. Amblard, and J. Mertz, Opt. Lett. 24, 969 (1999).

[5] J. P. Dunkers, M. T. Cicerone, and N. R. Washburn, Opt. Expr. 11, 3074 (2003).

[6] J. K. Barton, F. Guzman, and A. Tumlinson, J. Biomed. Opt. 9, 618 (2004).

[7] W. Denk, J.H. Strickler, and W.W. Webb, Science 248, 73 (1990).

[8] M. Rubart, Circ. Res. 95, 1154 (2004).

[9] J.A. Izatt, M.R. Hee, G. Owen, E.A. Swanson, and J.G. Fujimoto, Opt. Lett. 19, 590 (1994).

[10] A.D. Aguirre, P. Hsiung, T.H. Ko, I. Hartl, and J.G. Fujimoto, Opt. Lett. 28, 2064 (2003).

[11] R.A. Leitgeb, C.K. Hitzenberger, and A.F. Fercher, Opt. Exp. 11, 889 (2003).

[12] T.S. Ralston, D.L. Marks, P.S. Carney, and S.A. Boppart, J. Opt. Soc. Amer. A, In press (2005).

FIGURE CAPTIONS

FIG. 1.

Experimental setup for the integrated microscope. BS, beam splitter; CCD, charge-coupled line-scan camera; CU, collimating unit; D, dichroic; DG, diffraction grating; HWP, half-wave plate; M, reference-arm mirror; OBJ, objective; OF, neutral density optical filter; PMT, photomultiplier tube; SF, spectral filter; SM, scanning mirrors; SMF, single mode fiber; VF, variable neutral density filter. .

FIG. 2.

(Color online) Integrated optical imaging. (a) OCM, (b) MPM, and (c) corresponding histology of smooth muscle tissue from a transgenic mouse expressing GFP (Image size: 600 x 600 microns).

FIG. 3.

(Color online) Fibroblast cell cultured on a microtextured substrate. Grey-scale background: OCM; Green channel: MPM signal from GFP-vinculin; Blue channel: MPM signal from nuclear dye. (Image size: 200 x 200 microns).

FIG. 4.

(Color online) Fibroblast cells cultured in a 3D Matrigel matrix. Red channel: OCM backscattered signal from matrix; Green channel: MPM signal from GFP-vinculin; Blue channel: MPM signal from nuclear dye. (Image size: 200 x 200 microns).

FIGURE 1

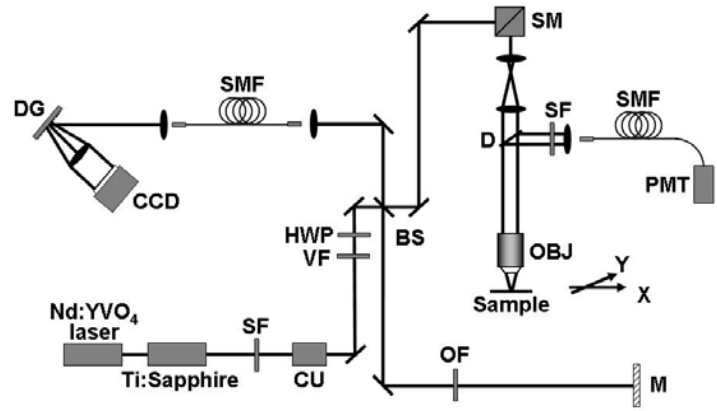

FIGURE 2

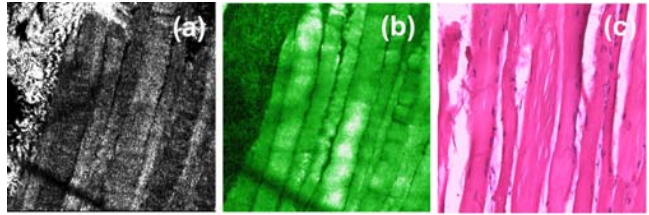

FIGURE 3

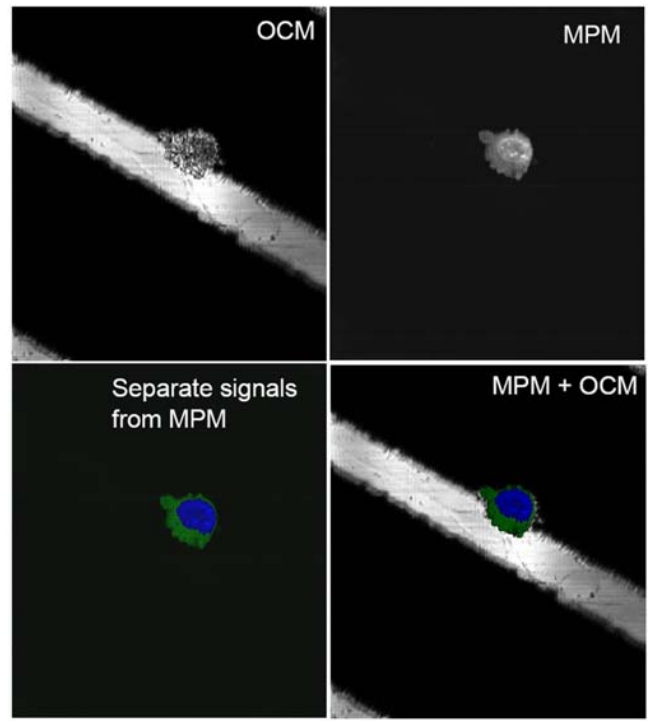

FIGURE 4

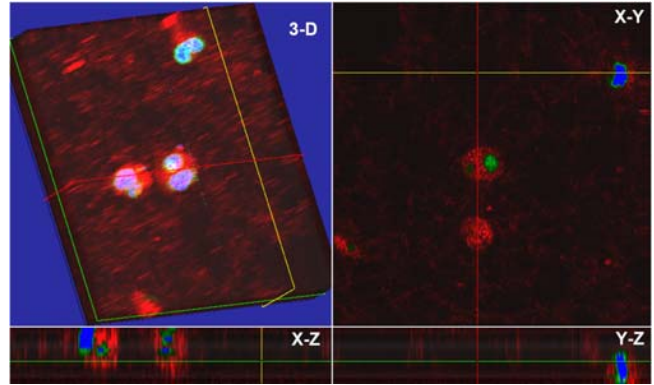